\documentclass[draft,jgrga]{AGUTeX}

\usepackage{graphicx}
\usepackage{lineno}
\usepackage{dirtytalk}
\usepackage{amsmath,xcolor}

\setkeys{Gin}{draft=false}

\authorrunninghead{ABDULLAH et al}

\titlerunninghead{Kinetic Ballooning Instability}

\begin{document}


\title{Evaluation of Kinetic Ballooning Instability in the Near-Earth Magnetotail}




\authors{Abdullah Khan \altaffilmark{1},
P. Zhu\altaffilmark{1,2} and A. Ali\altaffilmark{1,3}
\altaffilmark{}}
\altaffiltext{1}{CAS Key Laboratory of Geospace Environment and Department of Engineering and Applied Physics, University of Science and Technology of China, Hefei, Anhui 230026, China}

\altaffiltext{2}{Department of Engineering Physics, University of Wisconsin-Madison, Madison, Wisconsin 53706, USA}
\altaffiltext{3}{National Tokamak Fusion Program, Islamabad, 3329, Pakistan}


\begin{abstract}

Ballooning instabilities are widely believed to be a possible triggering mechanism for the onset of substorm and current disruption initiation in the near-Earth magnetotail. Yet the stability of the kinetic ballooning mode (KBM) in a global and realistic magnetotail configuration has not been well examined. In this paper, the growth rate of the KBM is calculated from analytical theory for the two-dimensional Voigt equilibrium within the framework of kinetic magnetohydrodynamic (MHD) model. The growth rate of the KBM is found to be strongly dependent on the field line stiffening factor $S$, which depends on the trapped electron dynamics, the finite ion gyroradius, and the magnetic drift motion of charged particles. Furthermore, calculations show that the KBM is unstable in a finite intermediate range of equatorial $\beta_{eq}$ values and the growth rate dependence on $\beta_{eq}$ is enhanced for larger $\rho_i$. The KBM stability is further analyzed in a broad range of $k_y$ for different values of ion Larmor radius $\rho_i$ and gradient ratio $\eta_j \equiv d\ln(T_j)/d\ln(n_j)$, where $T_j$ is the particle temperature and $n_j$ is the particle density. The KBM is found to be unstable for sufficiently high values of $k_y$, where the growth rate first increases to a maximum value and then decreases due to kinetic effects. The $k_y$ at the maximum growth rate decreases exponentially with $\rho_i$. The current sheet thinning is found to enhance the KBM growth rate and the unstable $\beta_{eq}$ regime in the near-Earth magnetotail.

\end{abstract}


\begin{article}

\section{Introduction}

The ballooning instability of the near-Earth magnetotail has long been suggested as a plausible trigger mechanism for the substorm onset. Several observational pieces of evidence support this idea. Based on the geostationary satellite (GEOS) 2 observational data, \citet {Roux et al 1991} suggested that the westward traveling surge observed by all-sky cameras was the image of the ballooning instability during a substorm onset in the equatorial region of the near-Earth magnetotail. Later, other GEOS 2 observational studies further elaborated the role of the ballooning instability as a trigger mechanism for the substorm onset [e.g. Pu et al, 1992, 1997]. Active Magnetospheric Particle Tracer Explorer/Charge Composition Explorer (AMPTE/CCE) satellite observation provides additional evidence that the ballooning instability can be a possible trigger for the substorm onset in the near-Earth magnetotail \citep {Cheng and Lui 1998}. In situ observations within the magnetotail configuration suggest that the current sheet breakup in the near-Earth before the substorm onset with no Earthward fast flow might be due to middle-magnetotail reconnection, whereas the Earthward fast flow might not trigger the substorm onset \citep {Erickson 2000,Ohtani 2002a,Ohtani 2002b}. \citet {Saito et al 2008} reported the Geotail observations which were consistent with the ballooning mode structures in the equatorial region of the near-Earth magnetotail prior to the substorm onset. Moreover, the observational studies also demonstrated that the predominant disruption modes prior to the substorm onsets have longitudinal wavelengths $\lambda_y$ during the expansion phase that peaks at $\sim 1000km$, corresponding to $k_y=2\pi/\lambda_y$ [e.g. Saito et al, 2008, Liang et al, 2009], where $k_y$ is in the direction of dawn-dusk. 

Numerous previous analytical and simulation studies regarding the investigation of ballooning instabilities in the near-Earth magnetotail have been based on the ideal MHD model. \citet {Hameiri et al 1991} suggested the existence of ballooning instabilities in the near-Earth magnetotail using the linear MHD theory. The ideal MHD studies indicated that the ballooning instability in the magnetotail would be stable due to strong plasma compression effect \citep {Lee and Wolf 1992,Ohtani and Tamao 1993,Lee and Min 1996}. The ideal MHD treatment of the ballooning mode by employing both eigenmode analysis and energy principle, results in stability conditions of different magnetotail configurations for the substorm onset \citep {Lee and Wolf 1992,Pu 1992,Lui et al 1992,Ohtani and Tamao 1993,Bhattacharjee 1998,Schindler and Birn 2004,Cheng and Zaharia 2004}. In all those calculations, the approximation $k_y\rightarrow\infty$ has been used. The ideal MHD simulation results of \citet {Bhattacharjee 1998} suggest that the ballooning instability may be a possible mechanism of the substorm onset. \citet {Wu et al 1998} and \citet {Zhu 2004} performed linear initial value MHD simulations to study the ballooning modes with finite $k_y$. Based on the Voigt equilibrium model [1986], \citet {Zhu 2004} investigated the finite $k_y$ ballooning instability. Their results showed that the growth rate of the ballooning mode increases with $k_y$ and gets saturated at large enough values of $k_y$. Nonideal and kinetic effects in the ballooning instability analyses were explored by using the drift and Hall MHD as well as the gyrokinetic models \citep {Pu 1992,Wu et al 1998,Cheng and Lui 1998,Ma and Bhattacharjee 1998,Lee 1999,Wong et al 2001,Zhu 2003,Zhu et al 2007,Crabtree et al 2003,Mazur et al 2013}. Although, the final outcome of stability analysis of the ballooning mode depends on the particular equilibrium model adopted for the near-Earth magnetotail, however, a common conclusion of all these studies is that the ballooning instability may possibly be responsible for triggering the substorm onset in the near-Earth magnetotail.

Ideal MHD model has long been used for the ballooning instability analysis. However, the ideal MHD based ballooning mode description does not take into account the kinetic effects. Some of the key deficiencies in the ideal MHD model originate from Ohm's law and adiabatic pressure law: (i) according to the ideal Ohm's law, the plasma is frozen to the magnetic field lines and moves across the fields with $\bf E \times \bf B$ drift velocity, thus ignoring the parallel electric field (ii) the plasma pressure is considered to vary according to the adiabatic pressure law. The Ohm's law and adiabatic pressure law may not be valid for many critical plasma phenomena. Therefore, for a more comprehensive description of the ballooning instability, it is indispensable to incorporate key kinetic effects into physical models. Kinetic-MHD is one such example which takes into account the dynamics of both trapped and untrapped electrons, the effects of finite Larmor radius (FLR), and the diamagnetic drift.

The kinetic MHD model has been implemented in several previous studies of the ballooning instability \citep {Cheng and Lui 1998,Wong et al 2001,Horton et al 1999,Horton et al 2001}. Kinetic effects such as trapped particle dynamics, FLR, wave-particle resonance and parallel electric field were shown to be important in determining the stability of KBM in the magnetosphere \citep {Cheng and Lui 1998}. Their theory is able to explicate the wave frequency, growth rate and high critical $\beta$ threshold of the low-frequency global instability observed by the AMPTE/CCE, where $\beta$ is the ratio of thermal pressure to the magnetic pressure. In particular, the trapped electron effect coupled with FLR effect produces a large parallel electric field that enhances parallel current and results in much higher stabilizing field line tension than predicted by the ideal MHD theory. As a result, a much higher critical value of $\beta$ than that predicted by the ideal MHD model was obtained. Furthermore, \citet {Horton et al 2001} investigated the stability of ballooning mode in geotail plasma and compared the calculation results between kinetic stability and MHD stability. \citet {Wong et al 2001} developed a general stability theory of kinetic MHD for drift modes at both low and high $\beta$ limits. These kinetic MHD studies provided a better understanding of the ballooning instability in wider parameter regimes of the near-Earth magnetotail plasmas. However, the stability of the KBM in a global and realistic configuration has been less studied.

This study is thus devoted to the evaluation of the KBM stability of a global near-Earth magnetotail configuration using the 2D Voigt equilibrium model. Previously, \citet {Zhu 2003,Zhu 2004} and \citet {Lee 1998,Lee 1999} have used the Voigt equilibrium model for investigating the ballooning mode stability of magnetotail configuration within the framework of ideal and Hall MHD models. In this paper, we study the stability of KBM in the global magnetotail plasmas based on the Voigt equilibrium model for the first time. Analysis shows that the KBM growth rate strongly depends on the parameter $S(=1+(n_e/n_{eu})|\delta|)$; where $n_e$ and $n_{eu}$ are the total and untrapped electron densities, respectively. The parameter $\delta$ depends on the untrapped/trapped electron dynamics, FLR effect and magnetic drift motion of the charged particles. The $k_y$ dependence of the KBM growth rate for different ion gyroradius $\rho_i$ and gradient ratio $\eta_j \equiv d\ln T_j/d\ln n_j$ are investigated in a broad range of $\beta_{eq}$ with different current sheet width of magnetotail, where $\beta_{eq}$ represent the value of $\beta$ at the equator. These results suggest that the excitation of KBM instability through current sheet thinning remains a viable scenario for substorm onset in the near-Earth magnetotail.

The rest of the paper is organized as follows. The Voigt equilibrium model for magnetotail plasma is presented in Section 2. In Section 3, we briefly recount the dispersion relation and the growth rate of the ballooning mode based on the kinetic MHD theory. In Section 4, we investigate the stability of the KBM in the magnetotail configurations using Voigt equilibrium model by evaluating the growth rate of ballooning mode from theory. Finally, a summary of the key results and a brief discussion on the new aspects of this study are presented in Section 5.


\section{Voigt Model of Magnetotail Equilibrium}

In the 2D Voigt equilibrium model, the magnetic field lies in $xz$ plane and can be expressed in terms of magnetic flux function $\Psi$ as $\bf B$ $=-\nabla \times (\Psi \hat y)=\hat y \times \nabla \Psi$ \citep {Voigt 1986}. Previously, this model has been applied in the ideal and Hall MHD studies of the ballooning instabilities in the magnetotail plasmas \citep {Zhu 2003,Zhu 2004}. In this study, we use the Voigt equilibrium model because it provides a rigorous formulation for the magnetostatic equilibrium. Since there is no X-point in the magnetic configuration, therefore, the role of pre-existing magnetic reconnection process may be eliminated. Moreover in the Voigt model, the equilibrium current density parallel to the magnetic field lines is kept zero to exclude the current driven instabilities. This is achieved in the Voigt equilibrium model by setting the magnetic field scaling parameter $h=0$. Therefore, in the Voigt equilibrium model, the only driving force for the ballooning instability in the bad curvature region of the magnetic field lines comes from the pressure gradient. Finally, the equilibrium current sheet thickness in the Voigt model could be varied by adjusting the equilibrium model parameters which allows the equilibrium magnetic field lines to vary from dipole-like to tail-like for a long range of plasma $\beta$. 

In this study, the geocentric solar magnetospheric (GSM) coordinates (x, y, z) are used, where $x$ is directed towards the sun, $y$ is from dawn-to-dusk, and $z$ is in the northward direction in the equatorial plane defined by the magnetic dipole axis of the Earth. For ballooning modes with large $k_y$, the near-Earth region of the magnetotail can be modeled using the 2D magnetostatic Voigt equilibrium \citep {Voigt 1986}. In this model, the Grad-Shafranov equation includes the 2D dipole magnetic field of the Earth and can be expressed as follows:
\begin{eqnarray}
\nabla^2 \Psi + \frac{d}{d\Psi}[P(\Psi) +\frac{1}{2}B_y^2(\Psi)]=-M_D \frac{\partial}{\partial x}\delta(x)\delta(z).
\end{eqnarray}
If we assume $P(\Psi)=\frac{k^{2}}{2}\Psi^{2}$ and $B_y(\Psi)=-h\Psi$, where $P$ is the equilibrium pressure and $B_y$ is the equilibrium magnetic field in the direction of dusk-dawn, $k$ and $h$ are the constant scaling parameters. Note that, in this study we assume $B_y=0$ by setting $h=0$. With these assumptions, equation (1) becomes linear in $\Psi$ and takes the following form
\begin{eqnarray}
\Psi(x,z)=-\frac{M_D}{2}\sum_{n=1}^{\infty}\cos(\eta_n z) e^{\lambda_n x} (1+e^{-2\lambda_n x_b})+\Psi_{-\infty}
\end{eqnarray}

\begin{figure}
\includegraphics[height=8.5 cm,width=15 cm]{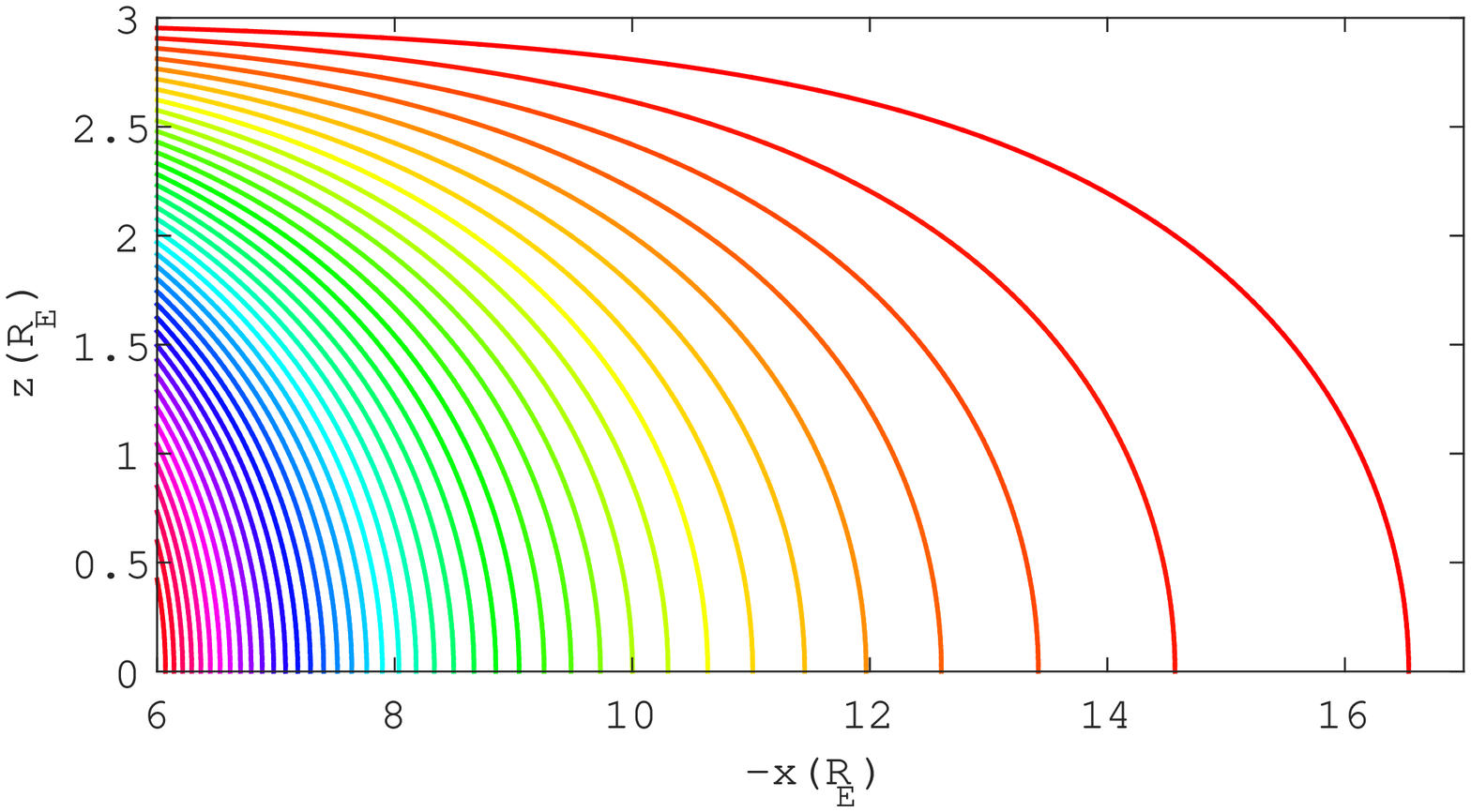}
 \caption{\label{pre_spit} Contour plot of magnetic flux function ($\Psi$) of the Voigt equilibrium in the near-Earth magnetotail. The equilibrium parameters are $x_b=6, z_{mp}=3, k^2=0.15$.}
\end{figure}
for the night-side magnetosphere $x<0$, where $M_D$ is the dipole moment of the Earth. The eigenvalues $\lambda_n$ and $\eta_n$ are related by two free physical parameters $h$ and $k$ as $\lambda_n^2=\eta_n^2-k^2-h^2$, where $\eta_n=({\pi}/{2})({2n-1})/z_{mp}$. The tail and day-side magnetopause locations are $z_{mp}$ and $x_b$ respectively. The magnetic flux function ($\Psi_{-\infty}$) at $x\rightarrow{-\infty}$ is considered zero in this study. The key parameters in the Voigt equilibrium model are $x_b$, $z_{mp}$, $k^2$ and $h$. The contour plot of the magnetic flux function $\Psi(x,z)$ of equation (2) is shown in Figure 1. For low values of the pressure parameter $k^2$, the magnetic field lines are round-shaped. As $k^2$ value is increased, the plasma $\beta_{eq}$ increases and the field lines become stretched as shown in Figure 1.

\section{Eigenmode Dispersion Relation of KBM}

The kinetic ballooning instability perturbations are considered in the regime $k_\perp \rho_i=O(1)$ and $k_\parallel\ll k_\perp$, where $k_\parallel$ and $k_\perp$ represent the parallel and perpendicular wave number, respectively. A local dispersion relation for the kinetic ballooning instability was derived for the frequency ordering $\omega\gg\omega_{de},\omega_{di}$. Neglecting the nonadiabatic density and pressure responses, the dispersion relation is obtained \citep {Cheng and Lui 1998}
\begin{eqnarray}
\frac{\omega(\omega-\omega_{\star pi})}{(1+b_i)V_A^2}\simeq {S}{k_{\parallel}^2}-\frac{2\kappa\cdot\nabla P}{B^2}.
\end{eqnarray}
Here $S$ is defined as follows \citep {Cheng and Gorelenkov 2004}:
\begin{eqnarray}
S&=& 1+\frac{n_e}{n_{eu}}\Big[\frac{\beta_e}{2}(\frac{\omega_{\star pi}-\omega_{\star pe}}{\omega})^2-\frac{q_i T_e}{q_e T_i}(\frac{\omega-\omega_{\star pi}}{\omega-\omega_{\star e}})b_i-\frac{3}{2}(\frac{\omega-\omega_{\star pe}}{\omega-\omega_{\star e}})\frac{<\hat\omega_{de}>}{\omega} \\ \nonumber
&+&(\frac{\omega-\omega_{\star pi}}{\omega-\omega_{\star e}})\frac{\hat\omega_{Be}+\hat\omega_{Ke}}{2\omega}\Big],
\end{eqnarray}\\
where $b_i=k_y^2\rho_i^2/2$, and $n_{eu}/n_e=1-(1-B(x_e)/B_{max})^{1/2}$ is the ratio of untrapped electron density to total electron density where $B(x_e)$ is the magnetic field at the equatorial points $(x_e)$ and $B_{max}$ is the maximum value of the magnetic field at the equator. The total electron density is $n_e=n_{eu}+n_{et}$ and the ratio of trapped electron density to total electron density is $n_{et}/n_e=(1-B(x_e)/B_{max})^{1/2}$. The untrapped electron fraction leads to the magnitude of stiffening factor $S$ being greater than unity. In equation (4), $\beta_e=2P_e/B^2$ represent the plasma parameter $\beta$ for electron and $\omega_{\star pj}=\omega_{\star j}(1+\eta_j)$ is the diamagnetic drift frequency due to pressure gradient for both ions and electrons, where $\omega_{\star j}={\bf B}\times\nabla P_j\cdot{\bf k_{\perp}}/(B\omega_{cj} P_j)({T_j}/{m_j})$ and $\eta_j=d\ln T_j/d\ln n_j$. Furthermore, $\hat\omega_{Be}$ and $\hat\omega_{Ke}$ represent the frequencies of electrons due to grad-B drift and curvature drift respectively. These are calculated as $\hat\omega_{Be}=2{\bf B}\times\nabla B\cdot{\bf k_{\perp}}T_e/q_e B^3$ and $\hat\omega_{Ke}=2{\bf B}\times{\kappa}\cdot{\bf k_{\perp}}T_e/q_e B^2$, where $\kappa=\hat{b}\cdot \nabla \hat{b}$ is the magnetic field curvature with $\hat{b}={\bf{B}}/B$, a unit vector along the Earth's magnetic field ${\bf{B}}$ with magnitude $B=|{\bf{B}}|$.

Thus, the linear growth rate from the local dispersion relation in equation (3) is
\begin{eqnarray}
\gamma_k^2 = (\frac{\beta_{eq}-S\beta^{MHD}}{L_pR_c})(1+b_i)V_A^2
\end{eqnarray}
where $R_c$ and $L_p$ are the radius of the magnetic field curvature and the pressure gradient scale length, respectively. $\beta_{eq}=2P/B^2=2P_i/B^2+2P_e/B^2$ is the equatorial $\beta$ where we assume $P_i=P_e$ ($P_i$ and $P_e$ represent the pressure of ions and electrons), $\beta^{MHD}=k_{\parallel}^2 L_pR_c$ is the predictable MHD $\beta$ based on the ideal MHD theory and $V_A=B/\sqrt{n_i m_i}$ is the Alfv\'en velocity. The mass ratio is considered to be $m_i/m_e=1836$.

\section{Numerical Results}

The key parameters in the KBM dispersion relation are evaluated as functions of $x_e$ in the simulation domain from $x_e=-6 R_E$ to $x_e=-16 R_E$ at $z=0$ for the Voigt equilibrium model. Figure 2 shows the variations of $L_p=P/(dP/dx)$, $\beta_{eq}$, and $R_c=1/|\hat{b}\cdot \nabla \hat{b}|$ start from the lowest values at $x_e=-6 R_E$ and approach to the highest values as $x_e\geq -9 R_E$. However, $\beta^{MHD}$ is maximum at $x_e=-6 R_E$ and decreases to minimum value for $x_e\geq -9 R_E$. We evaluate the KBM stability of the Voigt equilibrium in a broad range of $k_y$ for different values $\rho_i$ and $\eta_j$. Furthermore, the KBM stability is explored in a wide range of $\beta_{eq}$ for different values of $\rho_i$ and $z_{mp}$.

\begin{figure}[htbp]
\includegraphics[height=5.6 cm,width=7 cm]{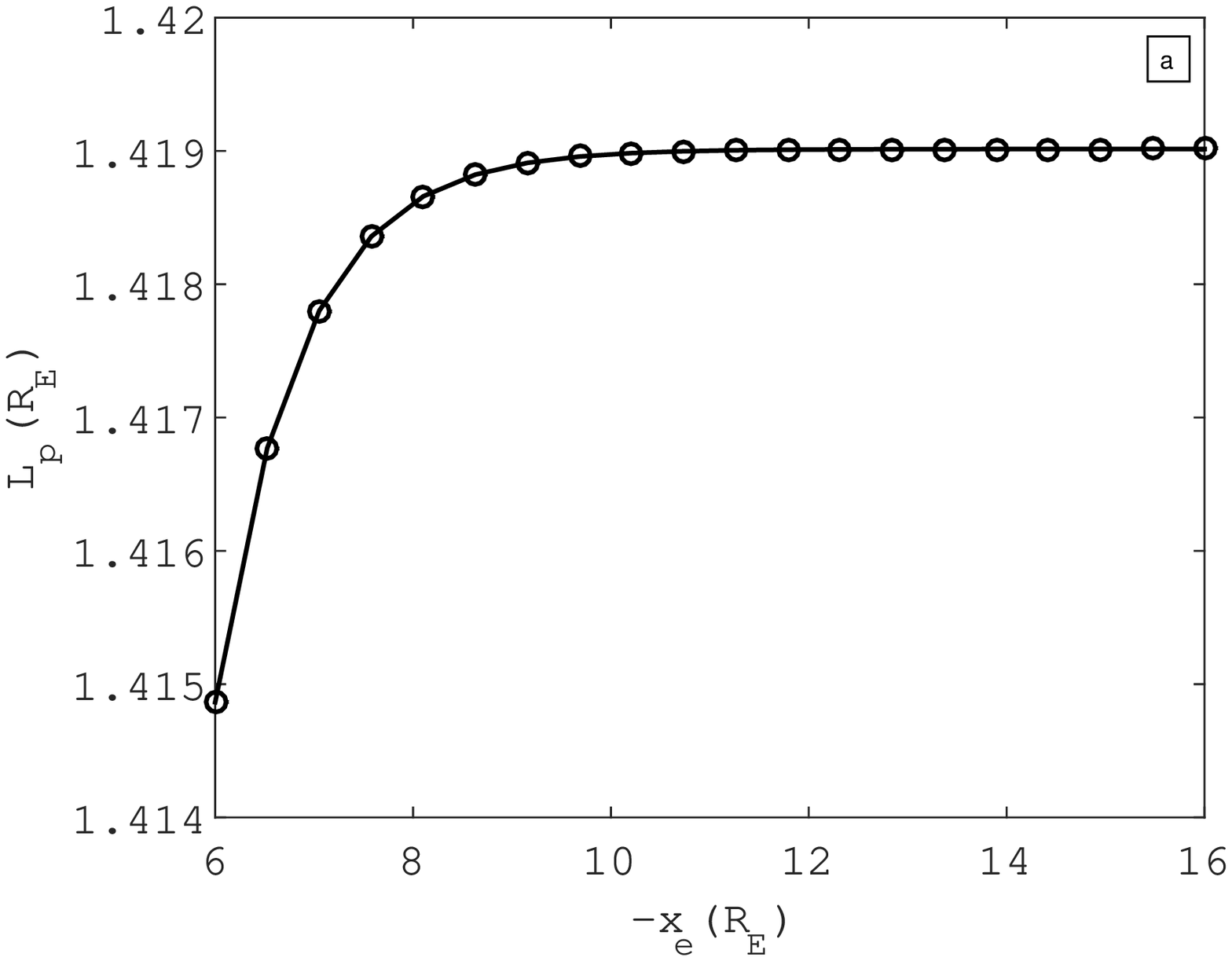}
\includegraphics[height=5.6 cm,width=7 cm]{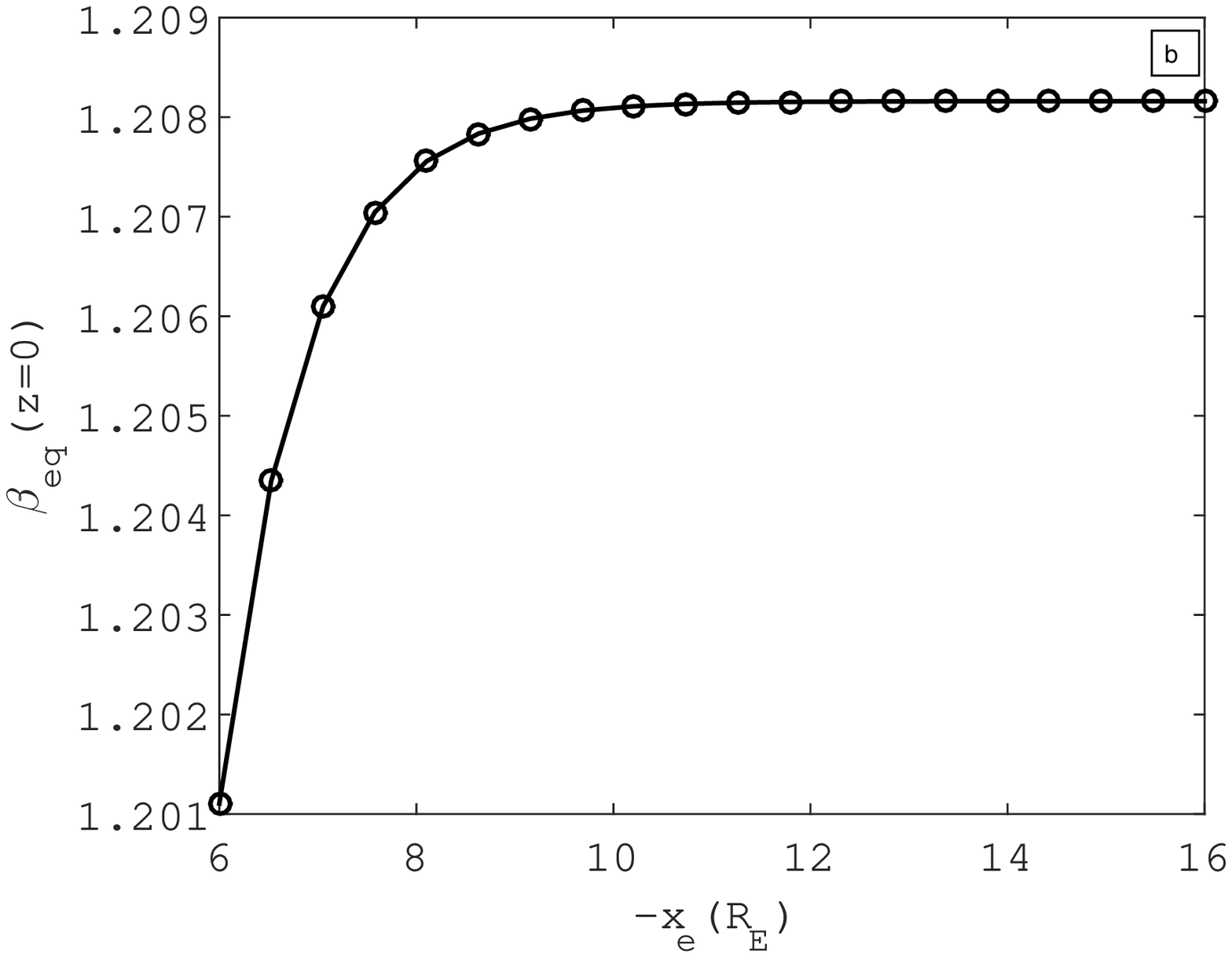}

\includegraphics[height=5.6 cm,width=7 cm]{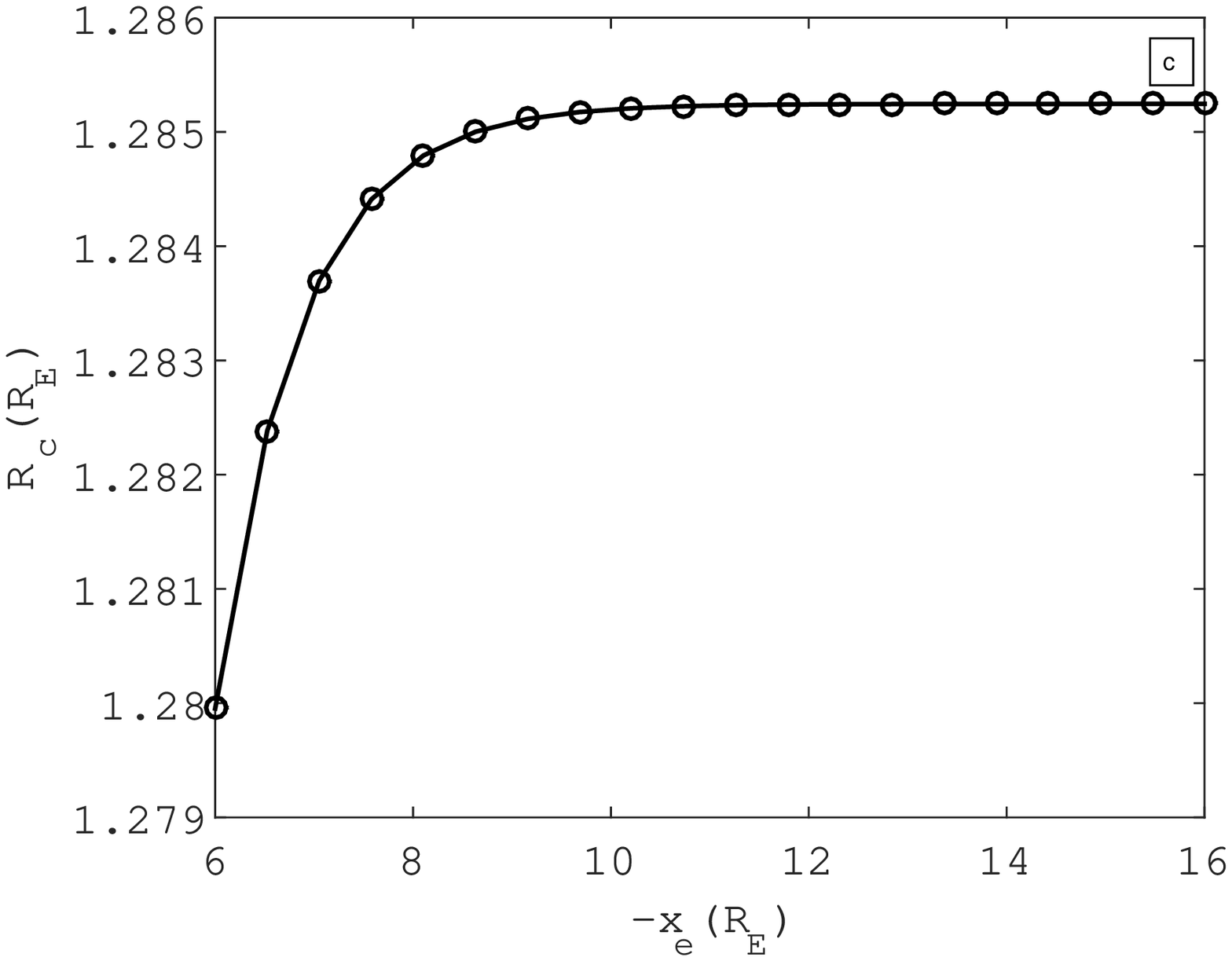}
\includegraphics[height=5.6 cm,width=7 cm]{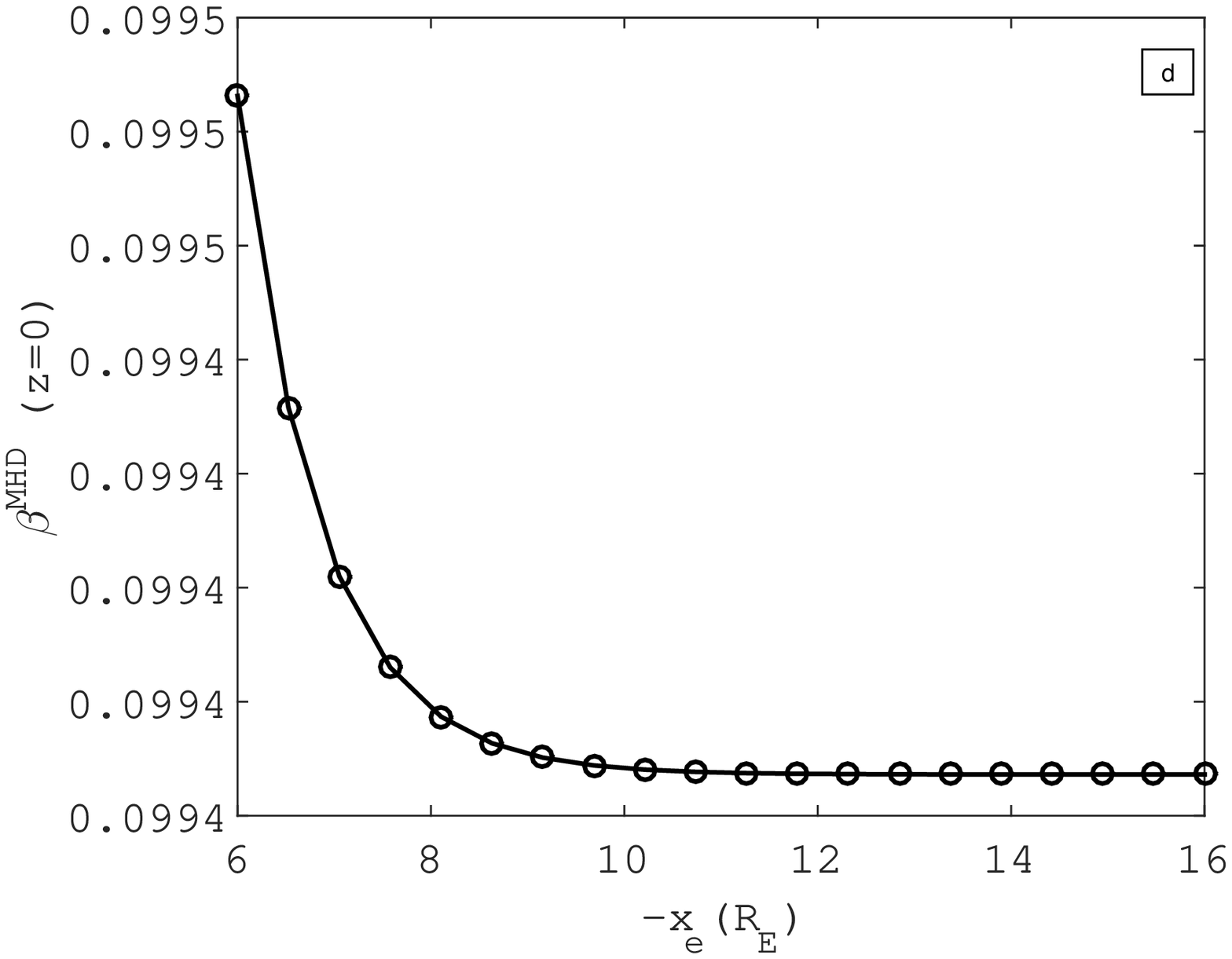}
\caption{\label{pre_spit}Variation of the key stability parameters as a function of $x_e$ at $z=0$ (a) the plasma scale length $L_p=P/(dP/dx)$, (b) the plasma parameter $\beta_{eq}=2P/B^2$, (c) the radius of curvature $R_c=1/|\hat{b}\cdot \nabla \hat{b}|$, and (d) $\beta^{MHD}=k_{\parallel}L_pR_c$. All these parameters are computed using the Voigt equilibrium model.}
\end{figure}

\subsection{$\beta_{eq}$  Dependence of KBM}

Next, we investigate the stability of KBM for high $\beta_{eq}$ plasmas in the near-Earth magnetotail. For the KBM analysis in the near-Earth magnetotail, the pressure scaling parameter ($k^2$) is used to control the local $\beta_{eq}$. However, the plasma $\beta_{eq}$ of the equilibrium increases as the $k^2$ increases and the magnetic field lines become more tail-like (stretched). The KBM growth rate as a function of $\beta_{eq}$ through the equatorial point $x_e=-9 R_E$ with $\eta_j=0$ is plotted for different values of $\rho_i=0.06, 0.07, 0.08, 0.09$ (in unit of Earth radius) as shown in Figure 3, where $k_y=4\pi/L_y$, $k_{\parallel}=\kappa/2$, $L_y=0.1 R_E$, and $T_e/T_i>1$. It is found that the onset of KBM at low end of $\beta_{eq}$ increase dramatically for different values of $\rho_i$. The KBM growth rate and the unstable regime of the ballooning mode increases significantly for larger values of $\rho_i$. Furthermore, the growth rate reaches the peak value and then decreases because the stiffening factor $S$ increases with $\beta_{eq}$. The stiffening factor plays a prominent role in the KBM stability of the near-Earth magnetotail.
\begin{figure}[htbp]
\includegraphics[height=8.5 cm,width=15 cm]{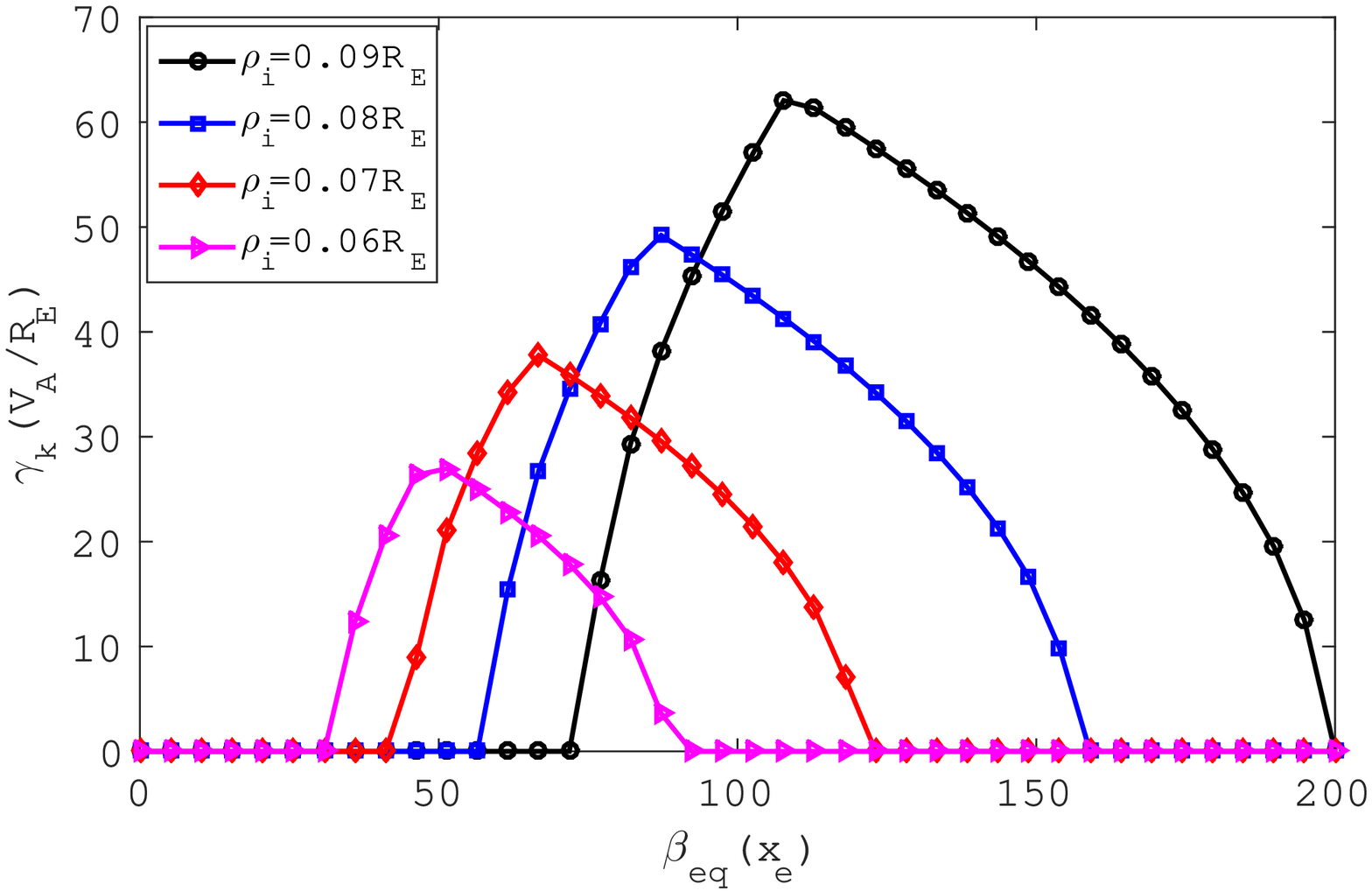}
\caption{\label{pre_spit}The KBM growth rate as a function of $\beta_{eq}$ for the wide current sheet configuration at the equatorial point $x_e=-9 R_E$ for four different values of $\rho_i$ at $\eta_j=0$.}
\end{figure}

\subsection{$k_y$  Dependence of KBM}

We evaluate the growth rate of KBM in a long range of perpendicular wave number $k_y$ using the Voigt equilibrium model at the Earth equator. In our simulations, the parallel wave number $k_{\parallel}$ along the field line is $\kappa/3$, where $\kappa$ is the magnitude of magnetic field curvature. In Figure 4, the KBM growth rate is plotted as a function of $k_y$ for different values of ion Larmor radius $(\rho_i)$ at $x_e=-9 R_E$ with $\beta_{eq}=1.208$. The value of $\rho_i$ can be evaluated as $\rho_i=v_{thi}/\omega_{ci}$, where $v_{thi}=(T_i/m_i)^{1/2}$ is the ion thermal velocity and $\omega_{ci}=q_iB/m_i$ is the ion gyrofrequency. The $\rho_i$ values is varied by changing the ion temperature. The KBM is found to be stable at low values of $k_y$ due to the stiffening factor $S$ as evident from Figure 4. The KBM becomes unstable for large values of $k_y$ where the onset of unstable regime depends on $\rho_i$. The KBM growth rate reaches the peak value and then decreases again due to stabilizing effects of both the FLR and the field line stiffening factor $S$. The $k_y$ at the maximum value of the growth rate is denoted by $k_y^{max}$ and it varies for different $\rho_i$ as shown in Figure 4. To elaborate this further, we plot $k_y^{max}$ as a function of $\rho_i$, as depicted in Figure 5. The $k_y^{max}$ decreases almost exponentially with $\rho_i$. To further elaborate the growth rate dependence on $k_y$, we consider the case of $\rho_i=0.07 R_E$. The stiffening factor $S$ decreases first and then increases with respect to $k_y$ as shown in Figure 6. This result is consistent with that of Figure 4. Such a stabilizing effect due to $S$ is unique to the near-Earth magnetotail configuration. There is also a weak dependence of KBM growth rate on the gradient ratio $\eta_j(\eta_j=d\ln T_j/d\ln n_j)$ shown in Figure 7.

 \begin{figure}[htbp]
\includegraphics[height=8.5 cm,width=15 cm]{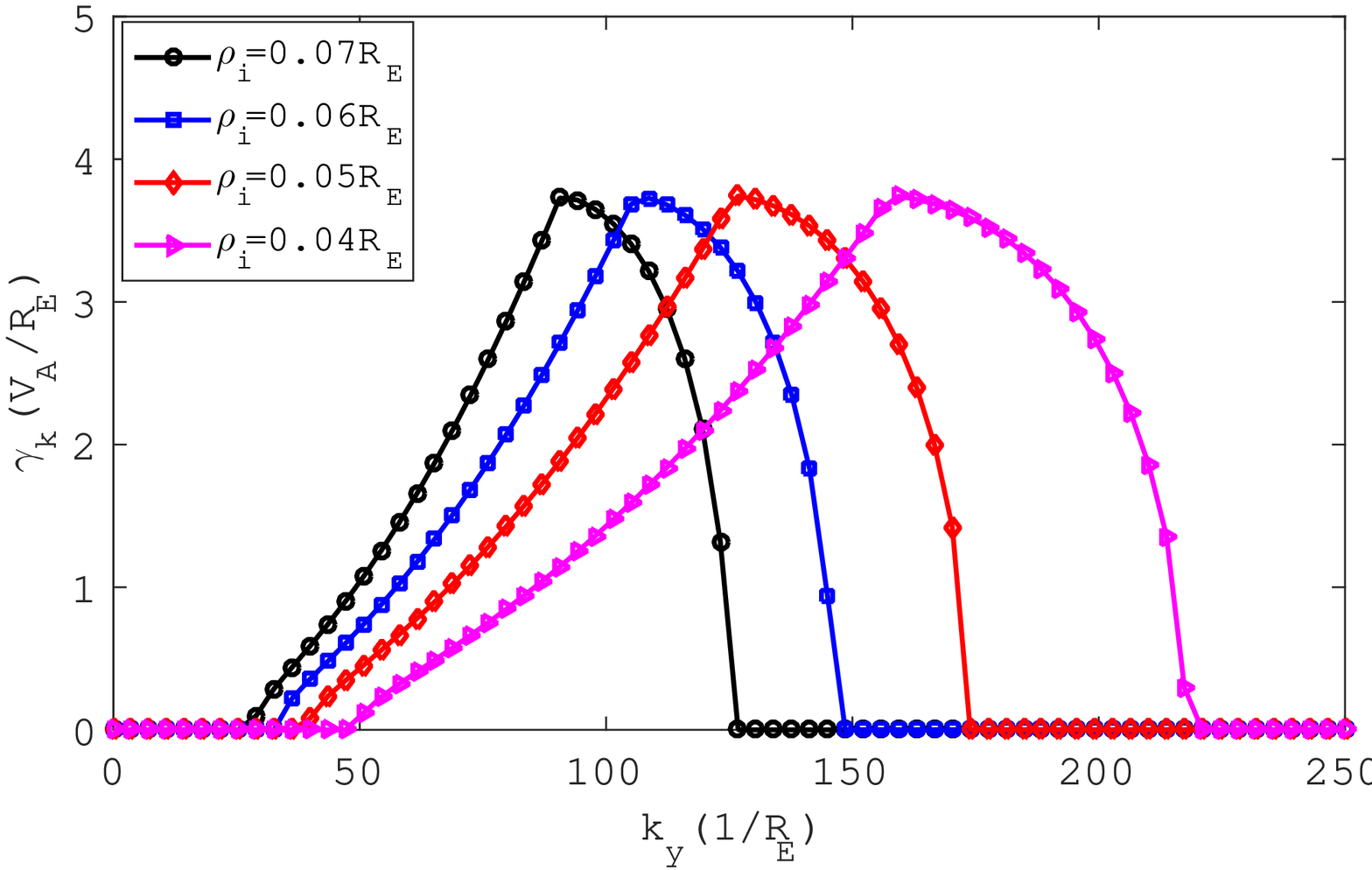}
\caption{\label{pre_spit} The KBM growth rate as a function of $k_y$ at the equatorial point $x_e=-9 R_E$ for four different values of $\rho_i$.}
\end{figure}

\begin{figure}[htbp]
\includegraphics[height=8.5 cm,width=15 cm]{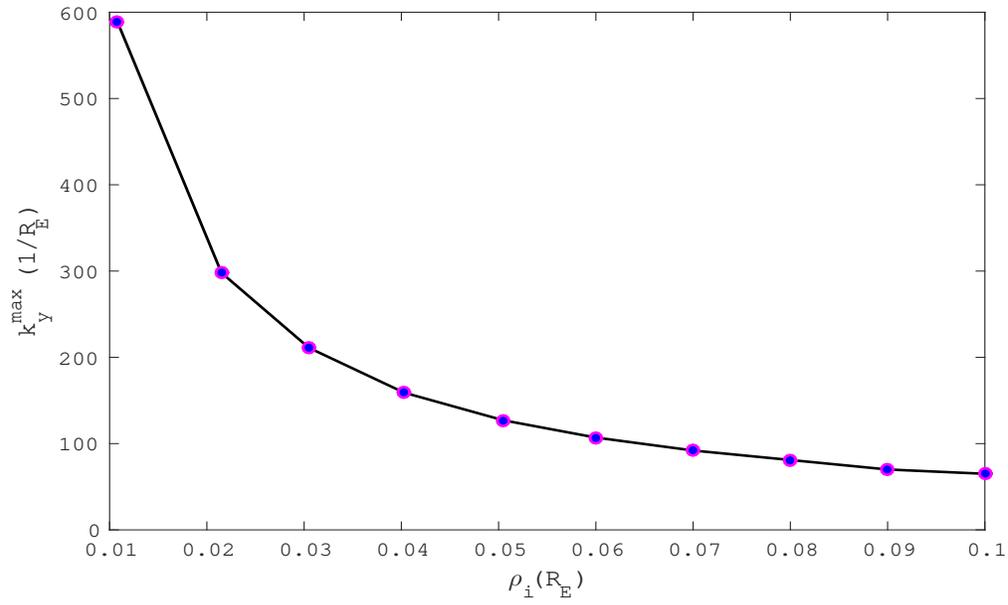}
\caption{\label{pre_spit}Variations of the maximum $k_y$ with $\rho_i$. }
\end{figure}

\begin{figure}[htbp]
\includegraphics[height=8.5 cm,width=15 cm]{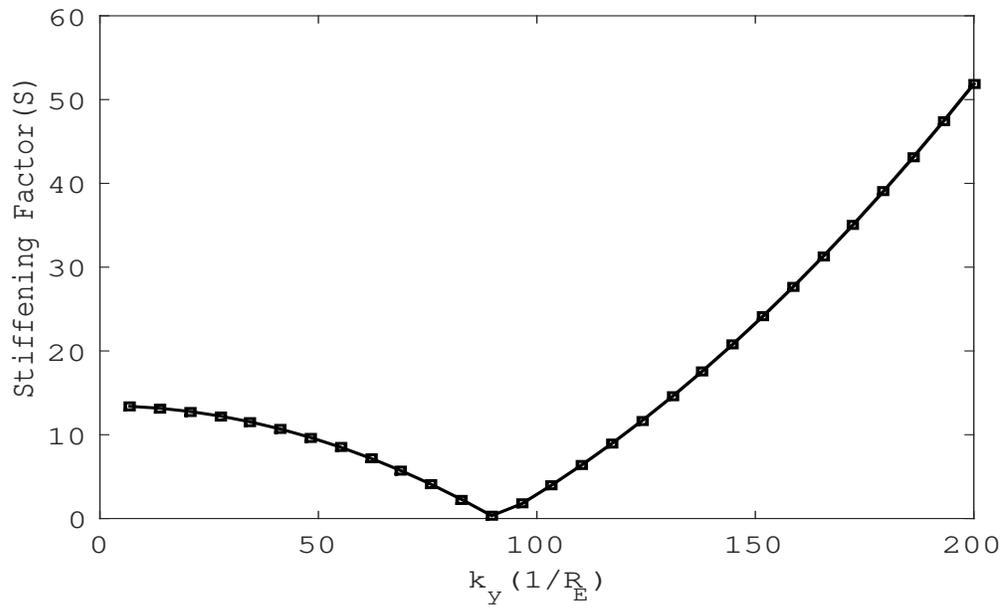}
\caption{\label{pre_spit}Variations of the stiffening factor $S$ in a long range of $k_y$.}
\end{figure}

\begin{figure}[htbp]
\includegraphics[height=8.5 cm,width=15 cm]{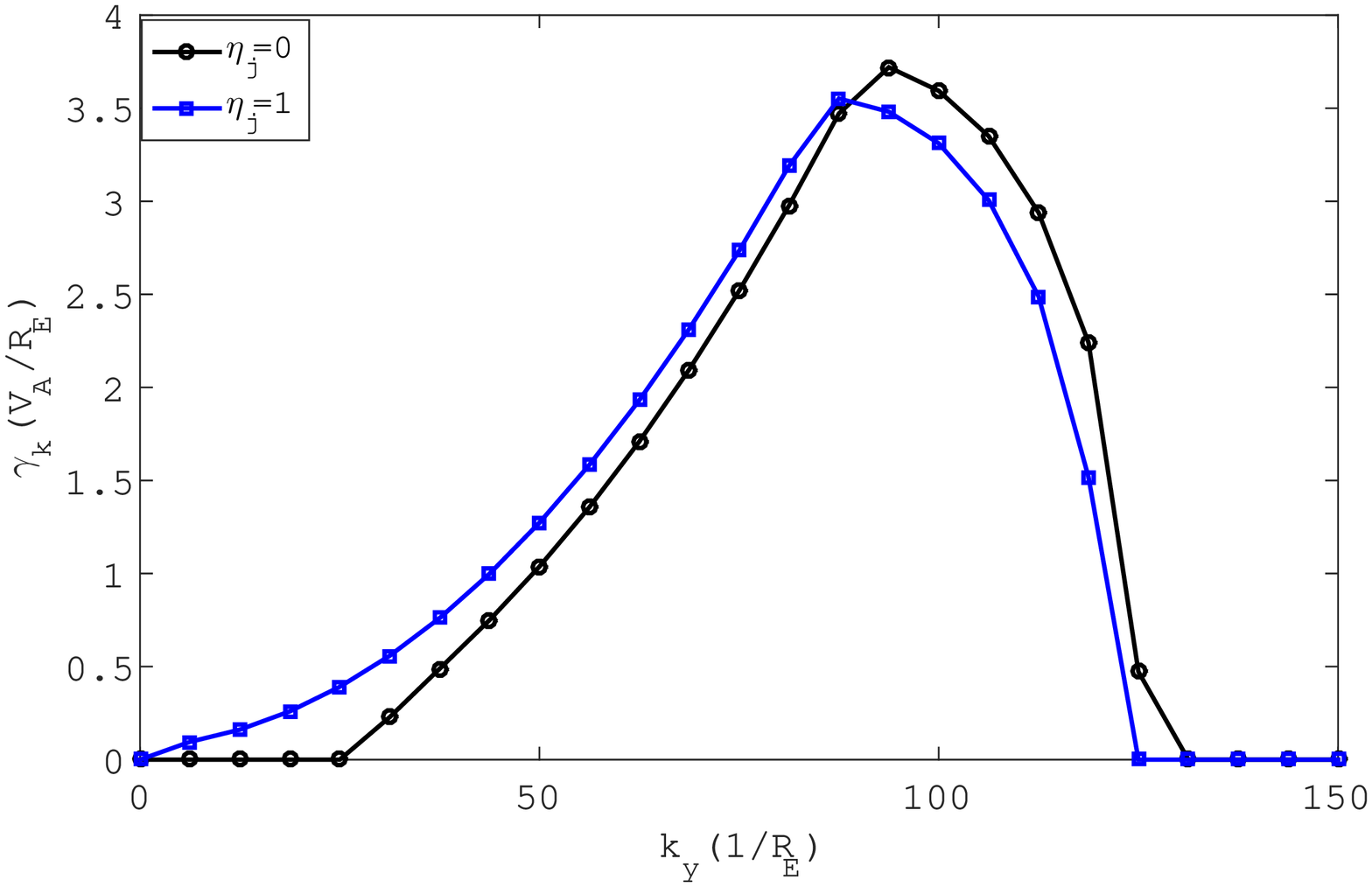}
\caption{\label{pre_spit}The KBM growth rate as a function of $k_y$ at the equatorial point $x_e=-9 R_E$ for two different values of $\eta_j$ at $\rho_i=0.07 R_E$. }
\end{figure}

We evaluate the maximum growth rate of the KBM in the near-Earth magnetotail as a function of $x_e$ using the Voigt equilibrium model (Figure 8). A specific case with $\rho_i=0.07 R_E$ and $k_y=92 (1/R_E)$ is selected, where the growth rate has a maximum value at $k_y=92$ as shown in Figure 4. The KBM is more unstable at $x_e=10.3 R_E$ in the near-Earth magnetotail for the Voigt equilibrium model. The KBM is stable at high values of $x_e$ at the equator due to the stabilizing factor of FLR effect and trapped electron dynamics.

\begin{figure}[htbp]
\includegraphics[height=8.5 cm,width=15 cm]{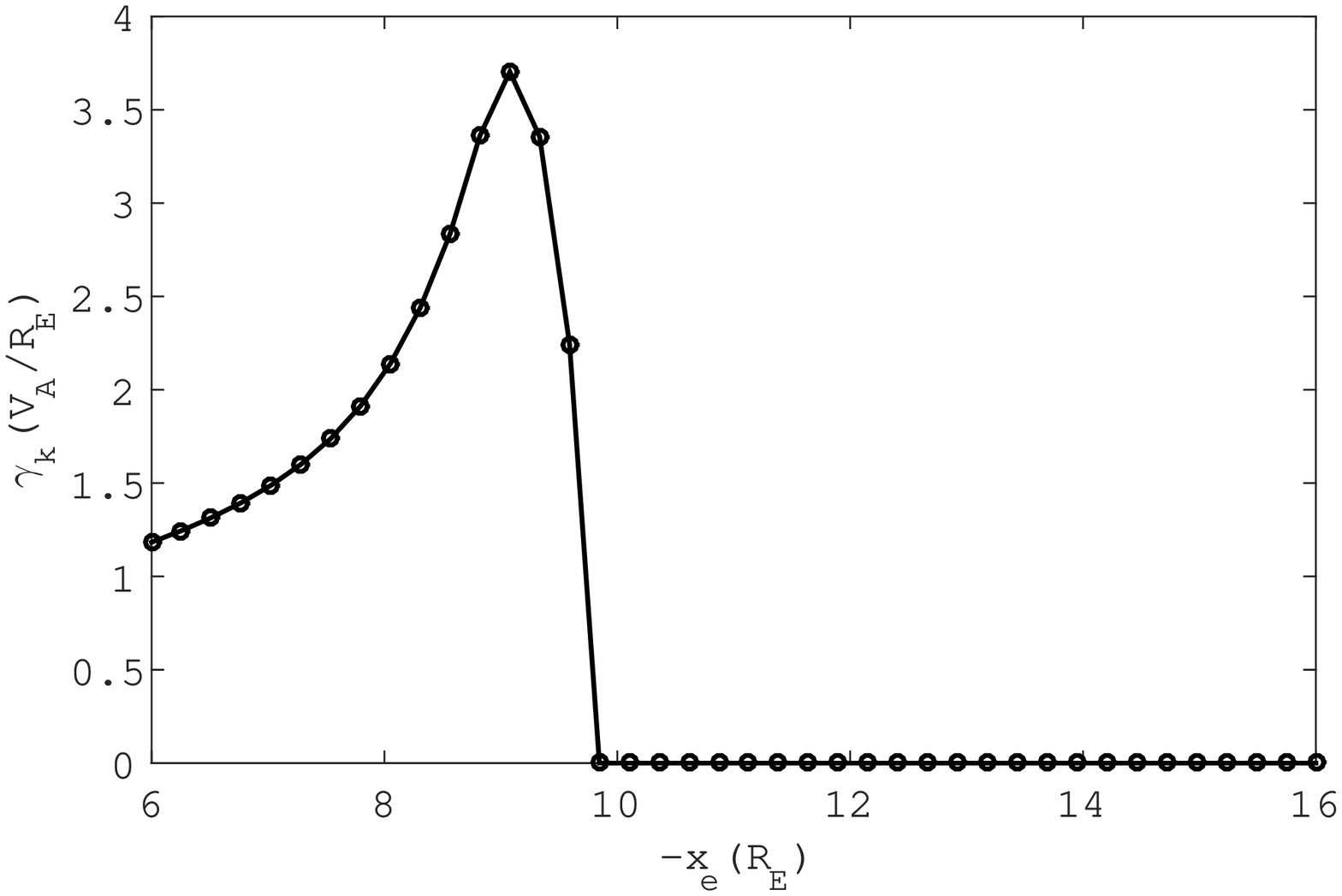}
\caption{\label{pre_spit}Maximum growth rate of the KBM as a function of $x_e$ at $z=0$ using the Voigt equilibrium model at $k_y=92 (1/R_E)$ and $\eta_j=0$.}
\end{figure}
\subsection {Thin Current Sheet Configuration of KBM}

Most of the thin current sheet studies \citep {Hau 1991,Becker 2001} investigate the current sheets in the tail region at $x_e=20 R_E$, and some simulation studies by \citet{Lee et al 1995,Erickson 1992} have considered the thin current sheet near the Earth at $x_e=-10 R_E$. Thinning of the current sheet is considered to play a critical role in the substorm onset dynamics. Thinning of the current sheet of the near-Earth magnetotail in the Voigt equilibrium model can be signified by reducing the value of $z_{mp}$ which is the distance between the equatorial plane and the night-side magnetopause location \citep {Zhu 2004}. We decrease the value of $z_{mp}$ from $3 R_E$ to $1 R_E$ in order to evaluate the KBM growth rate for finite value of $k_y$ for the case of $\rho_i=0.09 R_E$. We compare the $\beta_{eq}$ dependence of KBM growth rate for thin current sheet $(z_{mp}=1 R_E)$ and wider current sheet $(z_{mp}=3 R_E)$ (Figure 9). The KBM in the thin current sheet configuration is more unstable. This result suggests that the two possible scenarios proposed in ideal MHD model remain possible \citep {Zhu 2004}. In case of the wider current sheet, the KBM evolution may follow a path which is stable at low value of $\beta_{eq}$ and becomes unstable as the value of $\beta_{eq}$ exceeds the critical values. In case of the thin current sheet, the growth rate of the KBM may rapidly grow to the large amplitude and increases in a long range of $\beta_{eq}$ at the equator that may trigger the onset of substorm and current disruption.

\begin{figure}[htbp]
\includegraphics[height=8.5 cm,width=15 cm]{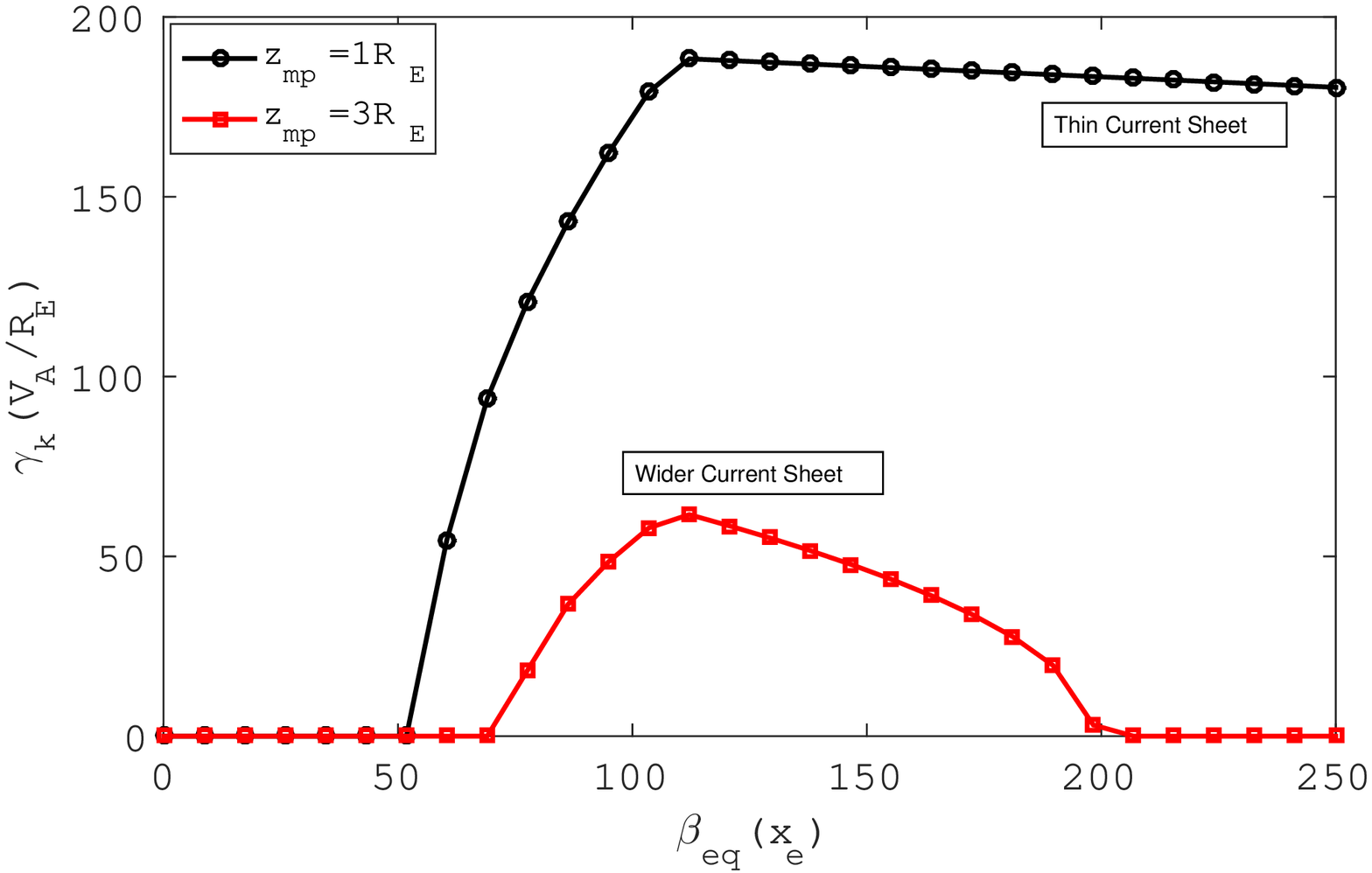}
\caption{\label{pre_spit}Comparison of the KBM growth rate for a thin $(z_{mp}=1 R_E)$ and a wide $(z_{mp}=3 R_E)$ current sheet configuration for specific value of $\rho_i$.}
\end{figure}

\section {Summary and Discussion}

In this paper, the KBM stability is comprehensively evaluated for the near-Earth magnetotail. Previous analytical theory reveals that the KBM growth rate changes significantly with the parameter $S$, which depends on the kinetic effects. The growth rate of KBM is numerically calculated for the 2D Voigt equilibrium model. The KBM stability is investigated in a wide range of $\beta_{eq}$ for different $\rho_i$ values. It is observed that the critical $\beta_{eq}$ for the KBM onset is higher for larger values of $\rho_i$ above which KBM growth rate increases with $\rho_i$. The growth rate of the KBM reach the peak value and then decreases monotonically as $\beta_{eq}$ increases for different $\rho_i$ values in the near-Earth magnetotail. It is found that the KBM is stable at low values of $k_y$ because of the stabilizing factor $S$ and becomes unstable for larger values of $k_y$. In the unstable regime, the growth rate first increases and reaches to a peak value and then decreases because of the stabilizing effects from both FLR and the field line stiffening factor $S$. Our results also suggest that the thin current sheet configuration enhances the KBM growth rate and may be responsible for triggering the substorm onset in the near-Earth magnetotail. 

The global eigenvalue approach for the evaluation of KBM growth rates instead of the local approximation will be studied in future.  Another question to be studied in the future is the KBM stability in the generalized Harris sheet configuration which may be more relevant to the near-Earth magnetotail prior to the substorm onset.






%
%
%


\acknowledgments
This research was supported by the Natural Science Foundation of China Grant No. 41474143, the 100 Talent Program of Chinese Academy of Sciences, and the U.S. DOE Grant Nos. DE-FG02-86ER53218 and DE-FC02-08ER54975. The computational work used the XSEDE resources (U.S. NSF Grant No. ACI-1053575) provided by TACC under Grant No. TG-ATM070010, and the resources of NERSC, which is supported by the U.S. DOE under Contract No. DE-AC02-05CH11231. The author Abdullah Khan acknowledges University of Science and Technology of China for awarding the Chinese Government Scholarship for his Ph.D. study. The author A. Ali acknowledges the support of the State Administration of Foreign Experts Affairs--Foreign Talented Youth Introduction Plan under Grant No. WQ2017ZGKX065. 

\end{article}


%
%
%
%
%
%





\end{document}